\documentclass[preprint,12pt]{elsarticle}
\usepackage[margin=1in]{geometry}
\usepackage{varioref,exscale,latexsym,amsmath,amssymb}
\usepackage{graphicx}
\usepackage{cancel}
\usepackage{inputenc}
\usepackage{amssymb}
\usepackage{natbib}
\usepackage{color}
\usepackage{lineno,hyperref}
\usepackage{feynmp}
\usepackage{subcaption}
\usepackage[font=scriptsize]{caption}

\newcommand{\be}{\begin{equation}} 
\newcommand{\ee}{\end{equation}}  
\newcommand{\bea}{\begin{eqnarray}}  
\newcommand{\eea}{\end{eqnarray}}

\newcommand{\dd}[2]{\frac{d #1}{d #2}}

\newcommand{\half}{\frac{1}{2}}

\modulolinenumbers[5]

\journal{Physics Letters B}
\begin{document}

\begin{frontmatter}
\title{Gravitational Radiation Background from Boson Star Binaries}

\author{Djuna Croon}\ead{djuna.lize.croon@dartmouth.edu}
\address{Department of Physics and Astronomy, Dartmouth College,
  Hanover, NH 03755, USA}

\author{Marcelo Gleiser}\ead{mgleiser@dartmouth.edu}
\address{Department of Physics and Astronomy, Dartmouth College,
  Hanover, NH 03755, USA}
  
  \author{Sonali Mohapatra}\ead{s.mohapatra@sussex.ac.uk}
\address{Department of Physics and Astronomy, University of Sussex, Falmer, Brighton, BN1 9QH, UK}

\author{Chen Sun}\ead{chen.sun@dartmouth.edu}
\address{CAS Key Laboratory of Theoretical Physics, Institute of Theoretical Physics, Chinese Academy of Sciences, Beijing 100190, P. R. China}
\address{Department of Physics and Astronomy, Dartmouth College,
  Hanover, NH 03755, USA}

\date{\today}

\begin{abstract} We calculate the gravitational radiation background generated from boson star binaries formed in locally dense clusters with formation rate tracked by the regular star formation rate. We compute how the the frequency window in gravitational waves is affected by the boson field mass and repulsive self-coupling, anticipating constraints from EPTA and LISA. We also comment on the possible detectability of these binaries.
\end{abstract}


\end{frontmatter}


\section{Introduction}

The recent detection of gravitational waves (GW) by LIGO and VIRGO 
have opened up a new window for our understanding of the physical properties of the universe \cite{Abbott:2016blz}. Probing the energy density of the stochastic Gravitational Wave Background (GRB) formed by the superposition of a large number of individual gravitational wave merger events is a long term goal of 
the next generation of GW detectors. It is thus of great interest to investigate different potential sources of GRBs and how to distinguish between their potential observational signatures. In this letter, we compute the GRB of an important class of hypothetical objects, merging binaries of Exotic Compact Objects (ECOs) composed of self-interacting scalar field configurations known as boson stars (BSs). Such objects were first proposed in the late 1960s \cite{Ruffini:1969qy} and further studied in the 1980s and 1990s \cite{Gleiser:1988rq, Jetzer:1991jr, Liddle:1993ha, Schunck:2003kk}, but are now experiencing a revival due to their potential role as dark matter candidates \cite{Giudice:2016zpa} and as remnants of early universe physics \cite{Mielke:2000mh}. The gravitational wave production from individual events of the merger of two boson stars has been studied in \cite{Palenzuela:2007dm} and \cite{Palenzuela:2017kcg}, for example. A preliminary estimate of the GRB in boson-star binary mergers was given in \cite{Gleiser:1989mb}. 

The success of inflationary cosmology \cite{Guth:1980zm} and the discovery of the Higgs Boson \cite{Chatrchyan:2012xdj} \cite{Aad:2012tfa} have opened up the possibility that different self-interacting scalar fields might exist in nature. The presence of such fundamental scalar fields in the early universe, maybe in dark matter clusters, could have led to their condensation into self-gravitating compact objects \cite{Levkov:2016rkk,Kolb:1993zz, Seidel:1993zk}. It is quite remarkable that for a repulsive self-interaction $\lambda|\phi|^4$ and a scalar field mass $m$, such objects have masses $M_{\rm BS} \sim \sqrt{\lambda}M_{\rm Pl}^3/m^2$, which, for $m/\lambda^{1/4}\sim m_p$, where $m_p$ is the proton mass, are parametrically equivalent to the Chandrasekhar mass \cite{Colpi:1986ye}.

Indeed, even a free, massive scalar field can generate a
self-gravitating object, supported against gravitational collapse
solely by quantum uncertainty \cite{Ruffini:1969qy}. This distinguishes them from fermionic
compact objects such as neutron stars (NS) and white dwarfs, which are
prevented from collapse due to degeneracy pressure \cite{ShapiroTeukolsky}. Another key
difference, important observationally to distinguish the two classes
of compact objects, is that the simplest BSs do not radiate
electromagnetically. 

In $\Lambda$CDM cosmology, using certain FDM
models, the first star formation in the center of
spherically-symmetric dark matter mini-halos have been found to be
around $z \sim 20-30$ \cite{Hirano:2017bnu},
\cite{Hirano:2013lba}. Given the uncertainty in the properties of such
primordial scalar fields, and to provide a more general analysis, we
assume here that BSs were formed at a rate that tracks the regular star formation rate, in locally-dense dark matter clusters. We will thus adopt this initial range of redshifts as a benchmark for our analysis. Our results can be extended to arbitrarily large redshifts.

As with their fermionic counterparts, BSs have a critical maximum mass
against central density beyond which they are unstable to
gravitational collapse into black holes (BHs) \cite{Gleiser:1988rq,Gleiser:1988ih}. In this
paper, we treat the two stars in the binary BS system as having
the same maximum mass and radius, which leads to the two objects
having the same compactness, defined as $C=G_NM/R$. The GRB is
typically characterized by the dimensionless quantity
$\Omega_{\rm{GW}}(f)$, the contribution in gravitational radiation in units of the critical density in
a frequency window $f$ and $f+\delta f$ to the total energy-density of
the universe  in a Hubble time. By studying their gravitational
imprints, we hope to gain insight on the properties of these exotic
objects, expanding the results of \cite{Gleiser:1989mb} and bringing them closer to current and planned observations.

\section{Boson Star properties}
\label{sec:boson-star-prop}

\subsection{Isolated Boson Stars}
\label{sec:isolated-boson-stars}

Very light bosons could form a Bose-Einstein condensate (BEC) in the early or late universe through various mechanisms \cite{Levkov:2016rkk,Kolb:1993zz, Seidel:1993zk}. 
Such objects are macroscopic quantum states that are prevented from
collapsing gravitationally by the Heisenberg uncertainty principle in
the non-interacting \cite{Ruffini:1969qy} and attractive self-interaction case \cite{Levkov:2016rkk}, or, in another possibility, through a
repulsive self-interaction that could balance gravity's attraction \cite{Colpi:1986ye}.
In this Letter, we study an Einstein-Klein-Gordon system with the following Lagrangian,
\be \mathcal{L} = \sqrt{-g}\left [ |(\partial \phi)|^2 -  m^2 |\phi|^2 - \half \lambda |\phi|^4  \right ],\ee
\noindent
where $\phi$ is a complex scalar field carrying a global $U(1)$. Real scalar fields can also form gravitationally-bound states, but these are time-dependent and have different properties \cite{Seidel:1991zh}.
Colpi {\it et al} showed that the maximum mass of a spherically-symmetric BS with repulsive self-interaction is given by \cite{Colpi:1986ye}
\begin{align}
\label{maxmass}
M_{*}^{max} \sim \frac{0.22 \, M_p^2\,\alpha^{1/2}}{m} \approx
  \frac{0.06 \sqrt{\lambda } M_p^3}{m^2},
\end{align}
where the rescaled coupling $\alpha$ is defined as $\alpha \equiv  {\lambda  \, M_p^2}/{(4 \pi \,  m^2)}$.
For a boson star with a repulsive self-interaction, the radius can be
estimated to be
\begin{align}
 \label{SSradius}
R_* \sim \frac{\sqrt{\lambda }}{\sqrt{G_N} m^2}.
\end{align}
The compactness of boson stars is discussed in many references such
as \cite{AmaroSeoane:2010qx,Giudice:2016zpa}. We note that the
compactness and mass of the stars are especially relevant for binary
GW events. Different formation mechanisms have been discussed in Refs. \cite{Levkov:2016rkk,Kolb:1993zz, Seidel:1993zk}.
However, since we are focussing here on the gravitational
radiation background, we need not worry about specific formation
mechanisms that lead to highly compact BSs. We will assume they exist and compute their contribution to the GRW.
We also note that if one assumes
the complex scalar $\phi$ to be responsible for the dark matter
in the Bullet Cluster, Ref.~\cite{Fan:2016rda} shows that
the constraint on the dark matter cross section
\cite{Markevitch:2003at,Randall:2007ph,Springel:2007tu} can be translated into a
bound on the boson's self-coupling, because the relative velocity of the Bullet Cluster is higher than the
sound speed of the condensate. 
The translated  bound on the self-interaction strength is
\be\label{Bullet} \lambda \lesssim 10^{-11} \left( \frac{m}{\text{eV}}\right)^{3/2}. \ee
 We note in passing that Ref.~\cite{Fan:2016rda} shows that BEC
requires light scalars $m< 1\mathrm{eV}$. However, the bound is based
on the inter-particle spacing estimated from the average density of
dark matter in the Universe. Since in the absence of a fundamental theory the exact formation process
of boson stars remain unclear, we consider the possibility of
their formation due to a large local density fluctuation. Therefore, we
do not worry about the bound on the scalar
mass.
In what follows, we saturate the Bullet Cluster bound and parametrize
the boson star mass effectively as
\begin{align}
  \label{eq:boson-star-mass-eff}
  M_*
  & =
    x M_*^{max}
    =2.5 \times 10^9 \; x \left ( \frac{\mathrm{eV}}{m} \right )^{5/4} 
    M_{\odot}, 
\end{align}
where $x$ is the fraction between boson star mass and the maximum stable
mass, and the radius as
\begin{align}
  \label{eq:boson-star-radius-eff}
  R_*
  & =
    y \frac{\sqrt \lambda}{\sqrt G_N m^2 }
    =
    1.1 \times 10^7\;  y \left ( \frac{\mathrm{eV}}{m} \right )^{5/4}  R_\odot,
\end{align}
where $y$ is the fraction or multiple of the star radius from Eq.~\eqref{SSradius}.

\subsection{Boson Star Binaries}
\label{sec:boson-star-binaries}

\begin{figure}[t] 
  \centering
  \includegraphics[width=.78\textwidth]{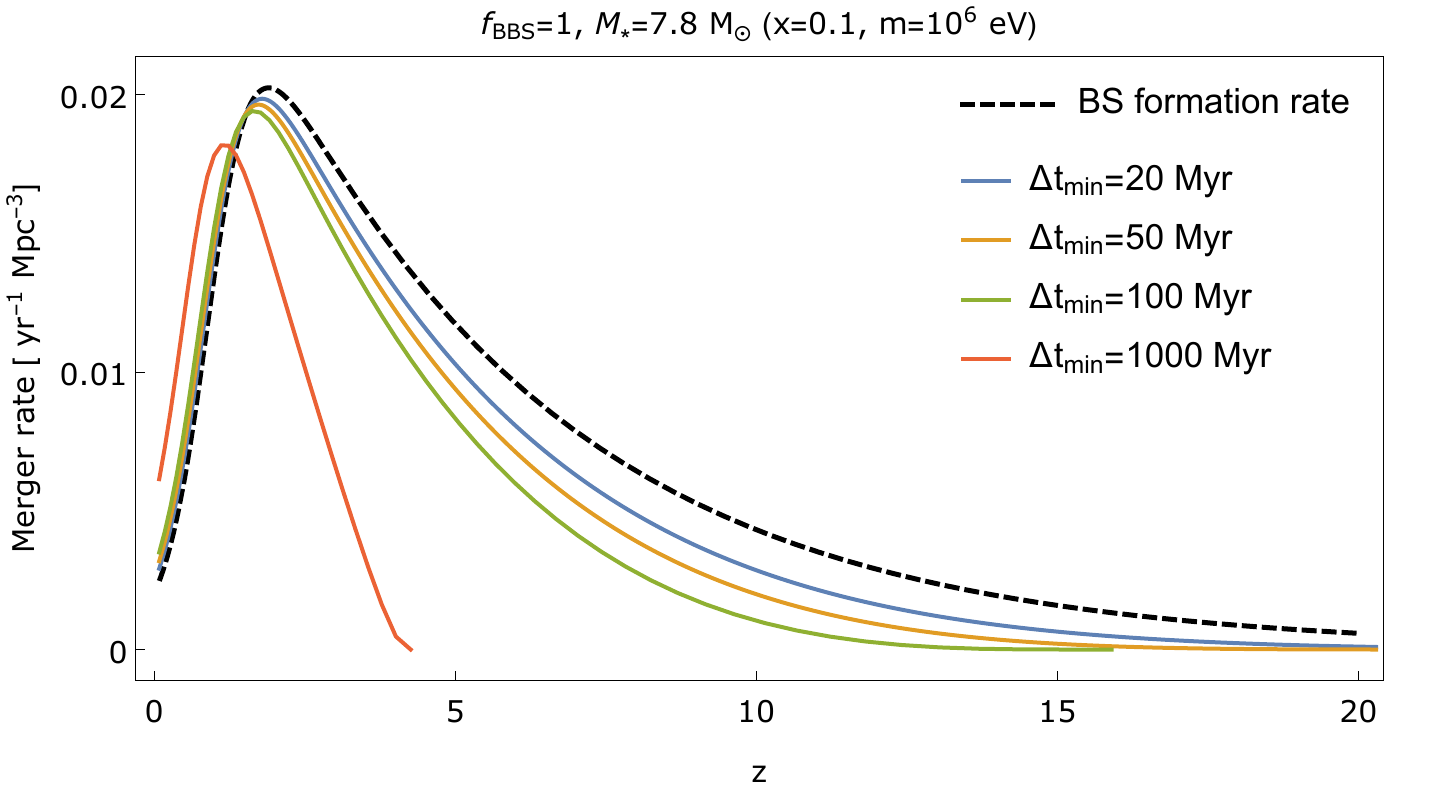}
  \caption{We take the shape of the regular star formation rate (dashed) from
    Ref.~\cite{Vangioni:2014axa} using the gamma ray burst fit therein
    ($\nu=0.16, z_m =1.9, a=2.76, b=2.56$), and assume that the
    boson star formation tracks the regular star formation with
    efficiency $f_\mathrm{BBS}$.
    We compare it with the
    merger rate (solid) calculated using Eq.~\eqref{eq:merger-rate}.
    It is observed that $\Delta t_{min}$,
    the minimum delay between formation and merging, has a small
    effect on the result 
    as long as the delay is comparable to NS mergers (20~Myr) \cite{Abbott:2017xzg} and BH mergers (50 Myr)
    \cite{TheLIGOScientific:2016wyq}. The
    benchmarks in Fig.~\ref{fig:GWplot} correspond to $\Delta t_{min}$ ranging from $10^{-12} \; \mathrm{Myr}$
    to $26 \; \mathrm{Myr}$.}
  \label{fig:bsrate-mergerrate}
\end{figure}

We briefly describe the properties of boson star
binaries that are relevant for the calculation of gravitational
radiation.
In what follows, we assume a conservative model for the estimation of
the binary formation rate, which tracks the star formation rate (SFR)
of luminous stars. Empirically, the luminous star-formation rate can be parametrized as a
function of redshift $z$  and stellar mass $M$ \cite{Springel:2002ux}, in units of
$\mathrm{yr}^{-1} \mathrm{Mpc}^{-3}$ as 
\begin{align}
\text{SFR}(z,M) = \text{SFR}_0 \, \left(\frac{M_\odot}{M\;\;}\right)  \frac{a\, e^{b(z-z_m)} }{a-b + b\,e^{a(z-z_m)} }\;.
\end{align}
The parameters $ \text{SFR}_0$, $z_m$, $a$, and $b$ are all determined
by fitting to observations such as gamma-ray burst rates and the galaxy
luminosity function. We adopt the fit from gamma-ray bursts from
\cite{Vangioni:2014axa}. We further parameterize the efficiency of the
binary boson star formation as a fraction of $\text{SFR}(z,M)$, denoted as
$f_{\text{BBS}}\leq 1$. The boson star binary
formation rate is, for a boson star of mass $M_*$ and formation redshift $z_f$,
\begin{align}
  \label{eq:BBSrate}
  R_{\mathrm{BBS}}(z_f, M_*) & = f_{\mathrm{BBS}} \times
                             \mathrm{SFR}(z_f, M_*).
\end{align}
Since we do not need all of the binaries to survive today to leave their
gravitational radiation imprint, we calculate the merger rate at redshift
$z$, which is mainly
determined by the binary formation rate at redshift $z_f$. On the
other hand, the larger the binary separation at formation, the less likely they
would have successfully merged, due to gravitational
perturbations from other sources. Following Ref.~\cite{Abbott:2017xzg}, we use an appropriately normalized weight function $p(\Delta t)$
to account for the merger efficiency, where $\Delta t $ is the time delay from formation of the binary to coalescence, 
\begin{align}
  \label{eq:merger-rate}
  R_m(t, M_*,f_\mathrm{BBS}) & =
                \int^{\Delta t_{max}}_{\Delta t_{min}} R_\mathrm{BBS}(t-\Delta t, M_*)
                \;
                p(\Delta t) \;d\Delta t.
\end{align}
Here, $\Delta t_{min}$ is the minimum time between formation and coalescence, and $\Delta t_{max}$ is determined by the maximum initial separation which allows for binary formation. As we will see below, the result is not sensitive to the precise choice of $\Delta t_{max}$. 
We will comment on a suitable $\Delta t_{min}$ for this integral in the following section.
We relate redshift to cosmic time with the approximate formula from Ref.~\cite{Carmeli:2005if},
\begin{align}
  \label{eq:cosmic-time}
t(z) &= \frac{2/H_0 }{1+ (z+1)^2},
\end{align}
\noindent
where $H_0$ is the Hubble constant today.
Next, let us estimate $p(\Delta t)$.
  For a pair of stars A and B, their initial separation $a$ defines a sphere
  inside which the number of stars is $N(a) =   \rho \pi a^3/6$. Assuming that the
  chance of any pair of stars forming a binary is roughly the same
  inside the sphere, the probability that stars A and B are bounded is
  \begin{align}
    \label{eq:counting-model}
    p(a) &= 
    \begin{pmatrix}
      N(a)\\
      2
    \end{pmatrix}^{-1}
    =
   \frac{2}{N(a) (N(a) -1)}
    \propto
    a^{-6}
   .
  \end{align}
This simple model captures the sharp decrease in the binary
  population as the pair separation increases. We
  note that the difficulty for binaries with initial large separation to
  form is not from perturbations that rip the two stars
  apart. Instead, the many `inbetweeners' are likely to
  form binaries with each of the two stars separately.
  Since gravitational radiation is the only channel for energy
  release, and since most of the initial binding and inspiraling process can be described by Newtonian
  dynamics, we use the merging time as in Ref. \cite{Croon:2017zcu}, 
  \begin{align}
    \label{eq:separation-to-time}
    \Delta t
    & \sim  a^4.
  \end{align}
This gives a weight function
  $p(\Delta t) \sim 1/\Delta t^{3/2}$.
\footnote{Note that this differs from
  Ref.~\cite{TheLIGOScientific:2016wyq,Abbott:2017xzg}, where a \textit{fiducial}
  model is used and the weight function for
NSs is chosen to be $p(\Delta t) \sim 1/\Delta t$. For a
study of different delay models, please refer to
Refs.~\cite{Tornatore:2007ds,madau2014cosmic,Mandel:2015qlu,TheLIGOScientific:2016htt}. }
This weight function also implies that the result is not
sensitive to $\Delta t_{max}$ and the precise determination of the initial separation.
The boson star formation rate and merger rate are shown in
Fig.~\ref{fig:bsrate-mergerrate}. As one can see, the merger rate is not very sensitive to $\Delta t_{min}$. The magnitude of the merger rate is controlled by $f_\mathrm{BBS}$,
which will be constrained together with their mass and radius.  

\section{Gravitational Waves from Boson Stars}

\subsection{Gravitational Waves from Single Binaries}
\label{sec:grav-wave-from-single}

The most important contribution to the stochastic background comes
from the inspiral phase of the binary mergers. In this stage, the
calculation can be done analytically. The system can be approximated
by a pair of purely self-gravitating point masses emitting mostly gravitational
quadrupole radiation. 
The radiation power is
\begin{align}
  P & = \frac{32}{5} {G_N \mu^2 \omega^6 r^4}.
\end{align}
Solving the dissipation equation $P = -\dot E$ gives us the
characteristic $f(t)\sim t^{-3/8}$ relation, and the radius as a
function of $t$, with $t$ being the time before coalescence,
\begin{align}
  \label{eq:quadrupole-radiation-power}
  f(t)
  & =
    \frac{5^{3/8}}{8\pi} (G_N m_c)^{-5/8} t^{-3/8},
    \cr
    r(t)
  & =
    \left ( \frac{256}{5} G_N^3 (M_A + M_B)M_A M_B \right )^{1/4}t^{1/4},
\end{align}
where $m_c$ is the chirp mass given by $m_c =  \frac{(M_A
  M_B)^{3/5}}{(M_A+M_B)^{1/5}}$, with $M_A, M_B$ being the masses of the
two stars. This approximation holds until the binary evolves beyond its innermost stable circular orbit (ISCO).
Inside the ISCO, tidal effects need to be taken into account, and the post-Newtonian expansion breaks down. 
The frequency of the ISCO is given by \cite{Giudice:2016zpa}
\begin{align} \label{fisco}
  f_\text{ISCO} = \frac{C_*^{3/2}}{3^{3/2}\, \pi \, G_N (M_1+M_2)},
\end{align}
which is a function of the compactness of the stars defined as
$ C_* \equiv {G_N M_*}/{R_* }$.
For boson stars with a fraction $x$ of the maximum mass \eqref{maxmass}, and a fraction or multiple $y$ of the radius \eqref{SSradius},
\begin{align}
  \label{eq:fISCO}
 f_\text{ISCO} \approx \frac{m^2 \sqrt{G_N}}{6 \sqrt{6} \pi
  ^{5/4}\,\sqrt{\lambda}} \, \sqrt{\frac{x}{y^{3}}} \approx 2.02 \times 10^{-15}  \; \text{Hz}\,
\sqrt{\frac{x}{y^{3}}} \, \sqrt{\frac{1}{\lambda }} \,
  \left(\frac{m}{\text{eV}} \right)^2 \,. 
\end{align}
If we saturate the Bullet Cluster bound as in Eq.~\ref{Bullet},
$f_\mathrm{ISCO}$ scales as $\sim m^{5/4}$.
\begin{align}
  \label{eq:fISCO-bullet-bound}
  f_\mathrm{ISCO}
  & \approx
    6.4\times 10^{-10}  \mathrm{Hz} \left (\sqrt{\frac{x}{y^3}}\right
    ) \left ( \frac{m}{\mathrm{eV}} \right )^{5/4}.  
\end{align}
We will estimate $\Delta t_{min}$ in \eqref{eq:merger-rate} based on the
following argument: if the boson star binary is formed at an initial
distance inside the ISCO, the binary will not experience an inspiral
phase. Therefore we choose $\Delta t_{min}$ to correspond to
$t_{ISCO}$, the time between entering the ISCO and
coalescence. %
In what follows, we sum up the contributions from
individual mergers to get the total gravitational radiation energy
density. When we do the summation, we use $f_{ISCO}$ as the cut off
frequency for each binary to guarantee the calculation based on
quadrupole radiation is valid. 

\subsection{Gravitational Radiation Energy Density}
\label{sec:grav-radi-energy-density}

\begin{figure}[htp]
   \centering
   \includegraphics[width=.78\textwidth]{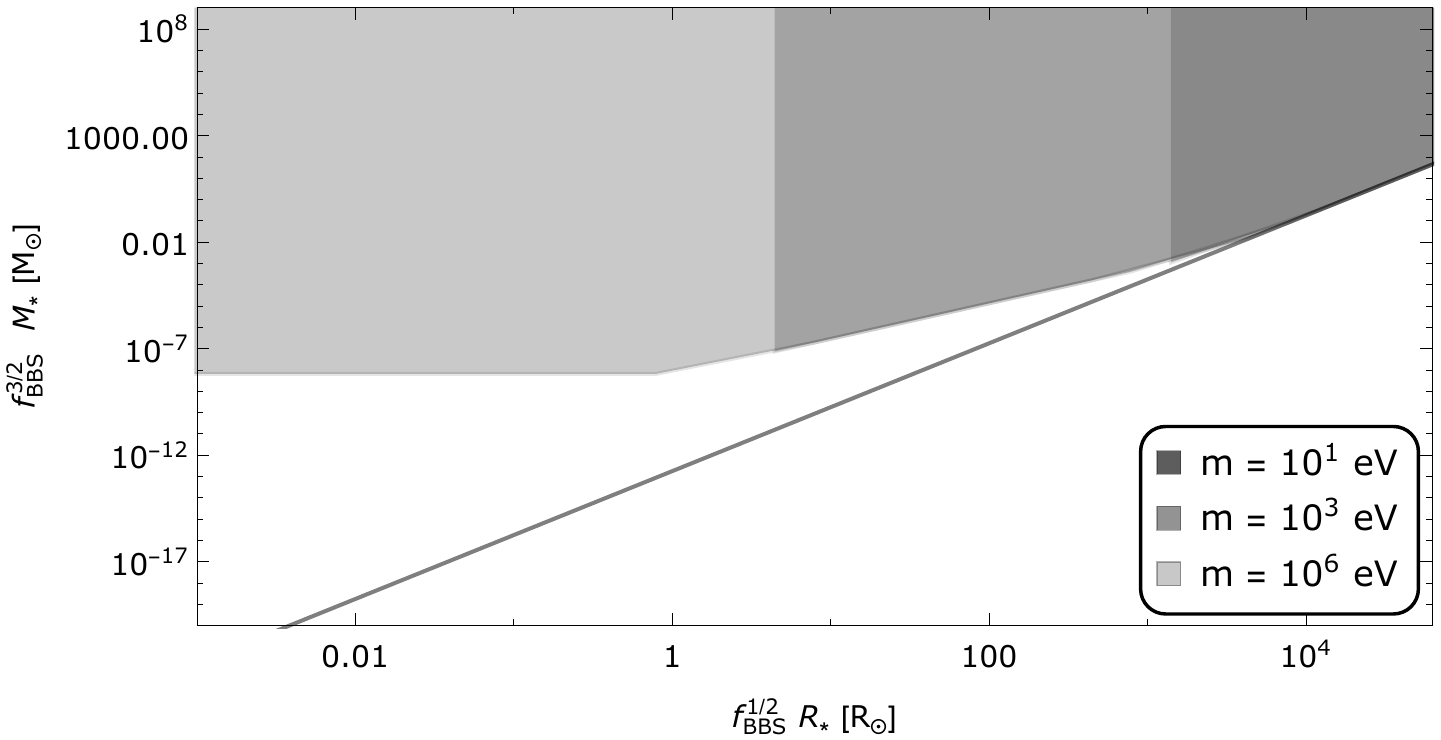}
   \caption{The bound on boson star parameters based on
     LISA. The gray region can be constrained by LISA. We take three
     benchmarks, with $m=10$ eV (the darkest region), $m=10^3$ eV (the
     both darker and the darkest region), $m=10^6$ eV (all colored
     area). The straight line is derived by setting $f_{ISCO} = 
     10^{-5} \; \mathrm{Hz}$, which is the lower end of LISA's
     sensitivity band. In this plot, LISA is not sensitive below the line.   
       }
   \label{fig:LISA-EPTA-bound} 
 \end{figure}

The energy spectrum of the gravitational radiation from boson stars is
defined as,
\be \Omega_\text{GW} (f) \equiv \frac{f}{\rho_c} \dd{\rho_{GW}}{f}, \ee
where $\rho_{GW}$ is the energy density of the gravitational wave in that frequency range and
$\rho_c$ is the critical energy density. 
Following  \cite{TheLIGOScientific:2016wyq}, this can be written using
the merger rate per unit of comoving volume per source time $R_m(z,
M_*)$, and the differential energy emitted by a single source $ d
E/df_s$ as, 
\begin{align}
 \label{masterOmega} 
  \Omega_\text{GW} (f, M_*, f_{BBS})
  &=
    \frac{f}{\rho_c H_0}
  \int^{z_{max}}_{0}
  \frac{R_m(z,M_*,f_\mathrm{BBS})}{(1+z)\sqrt{\Omega_M(1+z)^3+\Omega_\Lambda}} \, \frac{d
  E}{df_s}  \,   \, d z
  \cr
  &=
    {f^{2/3}}
    {f_\mathrm{BBS}\; \left (\frac{M_*}{M_\odot} \right )^{2/3}}  
    \left (
    \frac{\pi^{2/3}G_N ^{2/3} M_\odot^{5/3}}{ 2^{1/3}3\rho_c H_0} 
    \right )
    \int^{z_{max}}_{0}
    \frac{R_m(z,M_\odot,1)}{(1+z)^{4/3}\sqrt{\Omega_M(1+z)^3+\Omega_\Lambda}} \,
    \,   \, d z
    \cr
  & =
    1.91     \; f_\mathrm{BBS}\; x^{2/3}\;    \left ( \frac{f}{1\;\mathrm{Hz}} \right )^{2/3} \left
    (\frac{\mathrm{eV}}{m} \right )^{5/6}
,
\end{align}
\begin{figure}[t]
\centering
\includegraphics[width=0.88\textwidth]{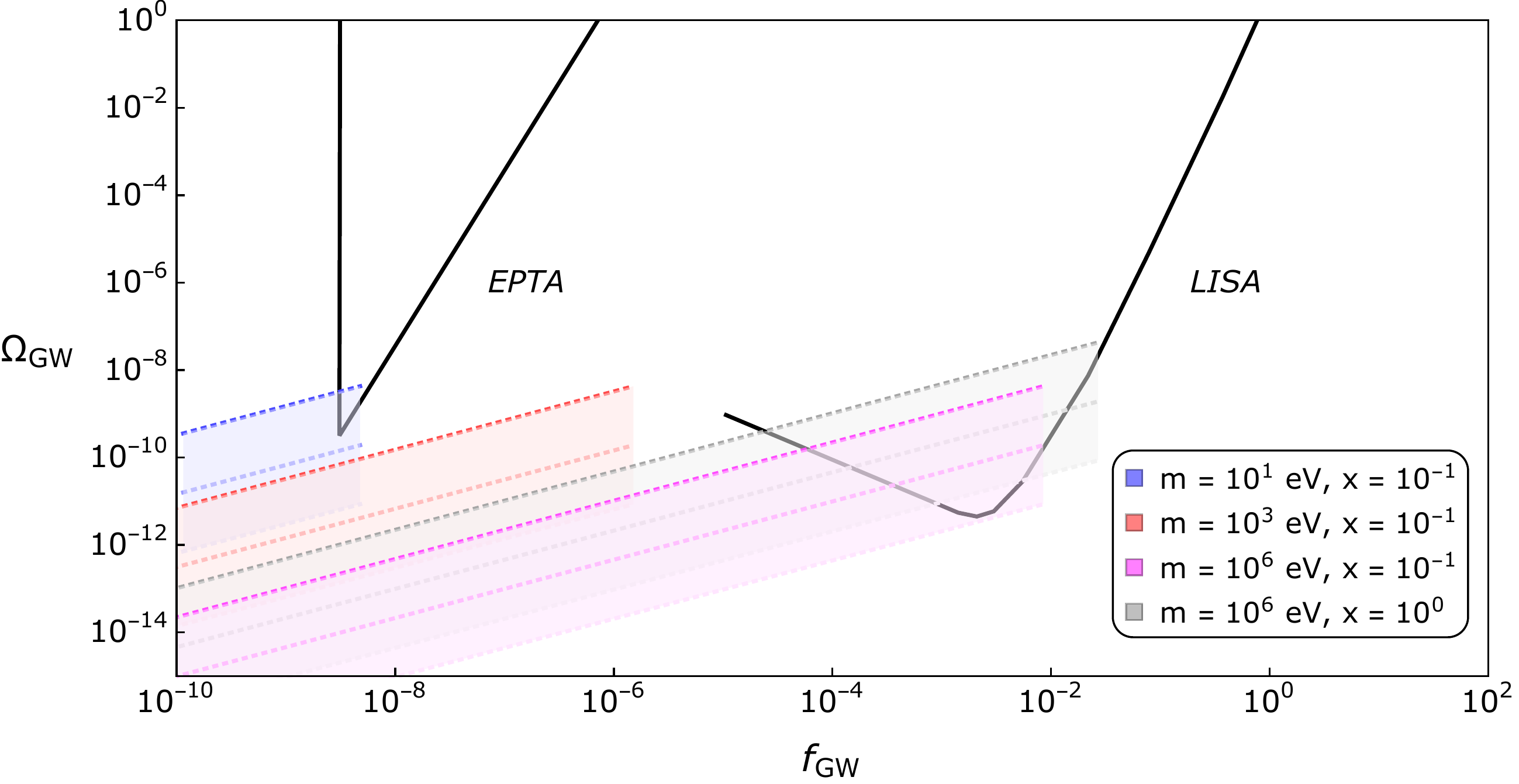}
\caption{
  \label{fig:GWplot}
  (Colored plot  online.)
  Plot of \eqref{masterOmega}. Here the fraction of the maximum boson star mass \eqref{maxmass} is taken conservatively to be $x = 10^{-1}$, and the fraction or multiple of the radius \eqref{SSradius} is taken as $y=1$. The self-coupling $\lambda$ has been chosen to saturate the Bullet Cluster constraint \eqref{Bullet}. 
The upper, lower, and middle lines are chosen for $f_{\text{BBS}} = 1/2$, $f_{\text{BBS}} = 10^{-3}$, and their geometric mean, respectively.  
Also shown are the EPTA \cite{Kramer:2013kea} and the LISA \cite{AmaroSeoane:2012km} exclusion prospects.}
\end{figure}
where we have used $f_s =  (1+z)f$ for the emitted (source) frequency,
and 
\begin{align}
  \label{eq:dEdfs}
  \frac{dE}{df_s}
  & =
    \frac{\pi^{2/3}}{3} {G_N ^{2/3} m_c^{5/3}}{f_s^{-1/3}}.
\end{align}
$f_\mathrm{ISCO}$ works as a cut-off at the high end of the
spectrum, which is shown in Eq.~\eqref{eq:fISCO-bullet-bound}. The spectrum is shown in Fig.\ref{fig:GWplot} for several benchmark scenarios.
In this plot, the fraction of the radius \eqref{SSradius} is taken as $y=1$.
 It is seen that the signal may be within reach of the next generation
 of gravitational wave interferometer experiments, and pulsar timing
 arrays. Also, we observe that the high end of the frequency band,
 determined by $f_\mathrm{ISCO}$, is proportional to $m^{5/4}$, if we
 saturate the Bullet Cluster bound, which indicates that boson stars
 consisting of heavy scalars are more likely to be probed by
 gravitational wave experiments.
 We show in Fig.~\ref{fig:LISA-EPTA-bound}
 the bound on binary formation efficiency
 $f_\mathrm{BBS}$, star mass, and star radius based on LISA for a few
 benchmarks of scalar mass. 

\section{Discussion}
As shown in Fig. \ref{fig:GWplot}, the gravitational signal from binaries of stars made of light bosons fall within the reach of the next generation of gravitational wave detectors and pulsar timing arrays. Failure to detect such spectra can be interpreted as a bound on the boson star parameters, as illustrated in Fig. \eqref{fig:LISA-EPTA-bound}. Such a bound can in turn be translated to bounds on the boson mass and self-coupling, once a specific formation scenario is assumed. 

The most important contribution to the boson star binary spectrum comes from the inspiral phase, which peaks at $f_\text{ISCO}$, the frequency corresponding to the innermost stable orbit. This peak frequency \eqref{fisco} is a function of the compactness of the boson stars, which depends on the scalar mass and self-coupling. This is to be compared with objects of which the compactness is known \cite{Giudice:2016zpa}. The compactness of a BH is 1/2, whereas realistic assumptions on the EOS for NSs would put them in the range $0.13 \lesssim C \lesssim 0.23$. For the boson stars considered here, the compactness saturates at $C\leq 0.16$, so close to the lower range of NSs and below that of BHs. We also note that BS mergers are not accompanied by electromagnetic signatures.

It is important to distinguish the stochastic background from boson stars from that due to more conventional binaries, such as BHs and NSs. Such a comparison relies on three main features. The stochastic spectrum is characterized by the fractional energy density $\Omega_\text{GW}(f)$ and the frequency band $f$. 
As is shown in equation \eqref{masterOmega}, $\Omega_\text{GW}(f)$ can be written as a function of the formation rate (parametrized by $f_\text{BBS}$) and the mass of the boson stars (as a function of $x$ and $m$). A fundamental difference is that boson star masses \eqref{maxmass} can take on a wide range of values, from that of NSs to that of supermassive BHs. Boson stars with a mass that falls outside the range typical for NSs and BHs are particularly interesting observationally. This corresponds to relatively heavy bosons, with $m \sim 10^5 \sqrt{x} \, \text{eV}$.
Also, a more exotic formation scenario than the one considered here may distinguish the boson star signal. For example, by considering redshifts different than the ones that track ordinary star formation. We leave the analysis of how these parameters impact the boson star stochastic background for future work.

 \section*{Acknowledgements}
 DC and CS would like to thank JiJi Fan for useful discussions.
 SM would also like to thank Xavier Calmet for useful feedback and
 discussions. MG is partially supported by a US Department of Energy grant DE-SC001038.
SM is supported by a Chancellor's International Research Scholarship of the University of Sussex and is grateful for the support of the TPP department at the University of Sussex. 
CS is supported in part by the International Postdoctoral
Fellowship funded by China Postdoctoral Science Foundation, and is
grateful for the hospitality and partial support of the Department of Physics and
Astronomy at Dartmouth College where this work was
done.


\begin{thebibliography}{999}%

\bibitem{Abbott:2016blz} 
  B.~P.~Abbott {\it et al.} [LIGO Scientific and Virgo Collaborations],
  Phys.\ Rev.\ Lett.\  {\bf 116}, no. 6, 061102 (2016)
  doi:10.1103/PhysRevLett.116.061102
  [arXiv:1602.03837 [gr-qc]].


\bibitem{Ruffini:1969qy} 
  R.~Ruffini and S.~Bonazzola,
  Phys.\ Rev.\  {\bf 187}, 1767 (1969).
  doi:10.1103/PhysRev.187.1767


\bibitem{Gleiser:1988rq} 
  M.~Gleiser,
  Phys.\ Rev.\ D {\bf 38}, 2376 (1988)
  Erratum: [Phys.\ Rev.\ D {\bf 39}, no. 4, 1257 (1989)].
  doi:10.1103/PhysRevD.38.2376, 10.1103/PhysRevD.39.1257


\bibitem{Jetzer:1991jr} 
  P.~Jetzer,
  Phys.\ Rept.\  {\bf 220}, 163 (1992).
  doi:10.1016/0370-1573(92)90123-H


\bibitem{Liddle:1993ha} 
  A.~R.~Liddle and M.~S.~Madsen,
  Int.\ J.\ Mod.\ Phys.\ D {\bf 1}, 101 (1992).
  doi:10.1142/S0218271892000057


\bibitem{Schunck:2003kk} 
  F.~E.~Schunck and E.~W.~Mielke,
  Class.\ Quant.\ Grav.\  {\bf 20}, R301 (2003)
  doi:10.1088/0264-9381/20/20/201
  [arXiv:0801.0307 [astro-ph]].


\bibitem{Giudice:2016zpa} 
  G.~F.~Giudice, M.~McCullough and A.~Urbano,
  JCAP {\bf 1610}, no. 10, 001 (2016)
  doi:10.1088/1475-7516/2016/10/001
  [arXiv:1605.01209 [hep-ph]].


\bibitem{Mielke:2000mh} 
  E.~W.~Mielke and F.~E.~Schunck,
  Nucl.\ Phys.\ B {\bf 564}, 185 (2000)
  doi:10.1016/S0550-3213(99)00492-7
  [gr-qc/0001061].


\bibitem{Palenzuela:2007dm} 
  C.~Palenzuela, L.~Lehner and S.~L.~Liebling,
  Phys.\ Rev.\ D {\bf 77}, 044036 (2008)
  doi:10.1103/PhysRevD.77.044036
  [arXiv:0706.2435 [gr-qc]].


\bibitem{Palenzuela:2017kcg} 
  C.~Palenzuela, P.~Pani, M.~Bezares, V.~Cardoso, L.~Lehner and S.~Liebling,
  Phys.\ Rev.\ D {\bf 96}, no. 10, 104058 (2017)
  doi:10.1103/PhysRevD.96.104058
  [arXiv:1710.09432 [gr-qc]].


\bibitem{Gleiser:1989mb} 
  M.~Gleiser,
  Phys.\ Rev.\ Lett.\  {\bf 63}, 1199 (1989).
  doi:10.1103/PhysRevLett.63.1199


\bibitem{Guth:1980zm} 
  A.~H.~Guth,
  Phys.\ Rev.\ D {\bf 23}, 347 (1981).
  doi:10.1103/PhysRevD.23.347


\bibitem{Chatrchyan:2012xdj} 
  S.~Chatrchyan {\it et al.} [CMS Collaboration],
  Phys.\ Lett.\ B {\bf 716}, 30 (2012)
  doi:10.1016/j.physletb.2012.08.021
  [arXiv:1207.7235 [hep-ex]].


\bibitem{Aad:2012tfa} 
  G.~Aad {\it et al.} [ATLAS Collaboration],
  Phys.\ Lett.\ B {\bf 716}, 1 (2012)
  doi:10.1016/j.physletb.2012.08.020
  [arXiv:1207.7214 [hep-ex]].


\bibitem{Levkov:2016rkk} 
  D.~G.~Levkov, A.~G.~Panin and I.~I.~Tkachev,
  Phys.\ Rev.\ Lett.\  {\bf 118}, no. 1, 011301 (2017)
  doi:10.1103/PhysRevLett.118.011301
  [arXiv:1609.03611 [astro-ph.CO]].


\bibitem{Kolb:1993zz} 
  E.~W.~Kolb and I.~I.~Tkachev,
  Phys.\ Rev.\ Lett.\  {\bf 71}, 3051 (1993)
  doi:10.1103/PhysRevLett.71.3051
  [hep-ph/9303313].


\bibitem{Seidel:1993zk} 
  E.~Seidel and W.~M.~Suen,
  Phys.\ Rev.\ Lett.\  {\bf 72}, 2516 (1994)
  doi:10.1103/PhysRevLett.72.2516
  [gr-qc/9309015].


\bibitem{Colpi:1986ye} 
  M.~Colpi, S.~L.~Shapiro and I.~Wasserman,
  Phys.\ Rev.\ Lett.\  {\bf 57}, 2485 (1986).
  doi:10.1103/PhysRevLett.57.2485


 \bibitem{ShapiroTeukolsky} S. L. Shapiro and S. A Teukolsky, {\it Black Holes, White Dwarfs, and Neutron Stars: The Physics of Compact Objects} (John Wiley \& Sons, New York NY 1983).
 
\bibitem{Hirano:2017bnu} 
  S.~Hirano, J.~M.~Sullivan and V.~Bromm,
  Mon.\ Not.\ Roy.\ Astron.\ Soc.\  {\bf 473}, no. 1, L6 (2018)
  doi:10.1093/mnrasl/slx146
  [arXiv:1706.00435 [astro-ph.CO]].


\bibitem{Hirano:2013lba} 
  S.~Hirano, T.~Hosokawa, N.~Yoshida, H.~Umeda, K.~Omukai, G.~Chiaki and H.~W.~Yorke,
  Astrophys.\ J.\  {\bf 781}, 60 (2014)
  doi:10.1088/0004-637X/781/2/60
  [arXiv:1308.4456 [astro-ph.CO]].


\bibitem{Gleiser:1988ih} 
  M.~Gleiser and R.~Watkins,
  Nucl.\ Phys.\ B {\bf 319}, 733 (1989).
  doi:10.1016/0550-3213(89)90627-5


\bibitem{Seidel:1991zh} 
  E.~Seidel and W.~M.~Suen,
  Phys.\ Rev.\ Lett.\  {\bf 66}, 1659 (1991).
  doi:10.1103/PhysRevLett.66.1659


\bibitem{AmaroSeoane:2010qx} 
  P.~Amaro-Seoane, J.~Barranco, A.~Bernal and L.~Rezzolla,
  JCAP {\bf 1011}, 002 (2010)
  doi:10.1088/1475-7516/2010/11/002
  [arXiv:1009.0019 [astro-ph.CO]].


\bibitem{Fan:2016rda} 
  J.~Fan,
  Phys.\ Dark Univ.\  {\bf 14}, 84 (2016)
  doi:10.1016/j.dark.2016.10.005
  [arXiv:1603.06580 [hep-ph]].


\bibitem{Markevitch:2003at} 
  M.~Markevitch {\it et al.},
  Astrophys.\ J.\  {\bf 606}, 819 (2004)
  doi:10.1086/383178
  [astro-ph/0309303].


\bibitem{Randall:2007ph} 
  S.~W.~Randall, M.~Markevitch, D.~Clowe, A.~H.~Gonzalez and M.~Bradac,
  Astrophys.\ J.\  {\bf 679}, 1173 (2008)
  doi:10.1086/587859
  [arXiv:0704.0261 [astro-ph]].


\bibitem{Springel:2007tu} 
  V.~Springel and G.~Farrar,
  Mon.\ Not.\ Roy.\ Astron.\ Soc.\  {\bf 380}, 911 (2007)
  doi:10.1111/j.1365-2966.2007.12159.x
  [astro-ph/0703232 [ASTRO-PH]].


\bibitem{Springel:2002ux} 
  V.~Springel and L.~Hernquist,
  Mon.\ Not.\ Roy.\ Astron.\ Soc.\  {\bf 339}, 312 (2003)
  doi:10.1046/j.1365-8711.2003.06207.x
  [astro-ph/0206395].


\bibitem{Vangioni:2014axa} 
  E.~Vangioni, K.~A.~Olive, T.~Prestegard, J.~Silk, P.~Petitjean and V.~Mandic,
  Mon.\ Not.\ Roy.\ Astron.\ Soc.\  {\bf 447}, 2575 (2015)
  doi:10.1093/mnras/stu2600
  [arXiv:1409.2462 [astro-ph.GA]].


\bibitem{Abbott:2017xzg} 
  B.~P.~Abbott {\it et al.} [LIGO Scientific and Virgo Collaborations],
  arXiv:1710.05837 [gr-qc].


\bibitem{Carmeli:2005if} 
  M.~Carmeli, J.~G.~Hartnett and F.~J.~Oliveira,
  Found.\ Phys.\ Lett.\  {\bf 19}, 277 (2006)
  doi:10.1007/s10702-006-0518-3
  [gr-qc/0506079].


\bibitem{Croon:2017zcu} 
  D.~Croon, A.~E.~Nelson, C.~Sun, D.~G.~E.~Walker and Z.~Z.~Xianyu,
  arXiv:1711.02096 [hep-ph].


\bibitem{TheLIGOScientific:2016wyq} 
  B.~P.~Abbott {\it et al.} [LIGO Scientific and Virgo Collaborations],
  Phys.\ Rev.\ Lett.\  {\bf 116}, no. 13, 131102 (2016)
  doi:10.1103/PhysRevLett.116.131102
  [arXiv:1602.03847 [gr-qc]].


\bibitem{Tornatore:2007ds} 
  L.~Tornatore, S.~Borgani, K.~Dolag and F.~Matteucci,
  Mon.\ Not.\ Roy.\ Astron.\ Soc.\  {\bf 382}, 1050 (2007)
  doi:10.1111/j.1365-2966.2007.12070.x
  [arXiv:0705.1921 [astro-ph]].


\bibitem{madau2014cosmic}
  P.~Madau, M.~Dickinson,
  Annual\ Review\ of\ Astronomy\ and\ Astrophysics {\bf 52}, 415 (2014)

\bibitem{Mandel:2015qlu} 
  I.~Mandel and S.~E.~de Mink,
  Mon.\ Not.\ Roy.\ Astron.\ Soc.\  {\bf 458}, no. 3, 2634 (2016)
  doi:10.1093/mnras/stw379
  [arXiv:1601.00007 [astro-ph.HE]].


\bibitem{TheLIGOScientific:2016htt} 
  B.~P.~Abbott {\it et al.} [LIGO Scientific and Virgo Collaborations],
  Astrophys.\ J.\  {\bf 818}, no. 2, L22 (2016)
  doi:10.3847/2041-8205/818/2/L22
  [arXiv:1602.03846 [astro-ph.HE]].


\bibitem{Kramer:2013kea} 
  M.~Kramer and D.~J.~Champion,
  Class.\ Quant.\ Grav.\  {\bf 30}, 224009 (2013).
  doi:10.1088/0264-9381/30/22/224009


\bibitem{AmaroSeoane:2012km} 
  P.~Amaro-Seoane {\it et al.},
  GW Notes {\bf 6}, 4 (2013)
  [arXiv:1201.3621 [astro-ph.CO]].

  
\end{thebibliography}
\end{document}